\documentclass[letterpaper,12pt]{article}
\usepackage[margin=1in]{geometry}
\usepackage{verbatim}
\usepackage{amsmath}
\usepackage{amssymb}
\usepackage{graphicx}
\usepackage{amsthm}
\usepackage{subfig}
\usepackage[round,authoryear]{natbib}

\def\epsilon{\varepsilon}

\def\phi{\varphi}

\def\0s{{\bf 0}}

\def\Prob{{\rm Prob}}

\def\s{\centerdot}

\newtheorem{theorem}{Theorem}[section]

\newtheorem{observe}[theorem]{Observation}
\newtheorem{remark1}[theorem]{Remark}

\newenvironment{remark}{\begin{remark1} \rm}{\end{remark1}}

\title{Testing the significance of assuming homogeneity
       in contingency-tables/cross-tabulations}
\author{Mark Tygert}

\begin{document}

\maketitle

\begin{abstract}
The model for homogeneity of proportions
in a two-way contingency-table/cross-tabulation is the same as the model
of independence,
except that the probabilistic process generating the data is viewed
as fixing the column totals (but not the row totals).
When gauging the consistency of observed data with the assumption
of independence,
recent work has illustrated that the Euclidean/Frobenius/Hilbert-Schmidt
distance is often far more statistically powerful than the classical statistics
such as $\chi^2$, the log--likelihood-ratio $G^2$,
the Freeman-Tukey/Hellinger distance, and other members
of the Cressie-Read power-divergence family.
The present paper indicates that the Euclidean/Frobenius/Hilbert-Schmidt
distance can be more powerful for gauging the consistency of observed data
with the assumption of homogeneity, too.

\medskip
\smallskip

\end{abstract}


\section{Introduction}
\label{intro}

\begin{table}[t]
\caption{A typical two-way contingency-table/cross-tabulation
($n_{j,k}$ is a nonnegative integer
with $j=1$,~$2$, \dots, $r$,\ \ \ $k=1$,~$2$, \dots, $s$;\ \ \
$n_{j,\s} = \sum_{k=1}^s n_{j,k}$ is a row total
with $j=1$,~$2$, \dots,~$r$;
$n_{\s,k} = \sum_{j=1}^r n_{j,k}$ is a column total
with $k=1$,~$2$, \dots, $s$;
and $n_{\s,\s} = \sum_{j=1}^r \sum_{k=1}^s n_{j,k} = n$ is the grand total)}
\label{tab}
\vspace{-2.5em}
\begin{center}
$$
\begin{array}{c|cccc|c}
       &       1 &       2 & \cdots &       s & \\\hline
     1 & n_{1,1} & n_{1,2} & \cdots & n_{1,s} & n_{1,\s} \\
     2 & n_{2,1} & n_{2,2} & \cdots & n_{2,s} & n_{2,\s} \\
\vdots &  \vdots &  \vdots & \ddots &  \vdots &  \vdots \\
     r & n_{r,1} & n_{r,2} & \cdots & n_{r,s} & n_{r,\s} \\\hline
       & n_{\s,1} & n_{\s,2} & \cdots & n_{\s,s} & n_{\s,\s}
\end{array}
$$
\end{center}
\vspace{1em}
\end{table}

\begin{table}
\caption{The model for homogeneity of proportions
($n_{1,\s}$, $n_{2,\s}$, \dots, $n_{r,\s}$ are the row totals;
$n_{\s,1}$, $n_{\s,2}$, \dots, $n_{\s,s}$ are the column totals;
and $n_{\s,\s} = n$ is the grand total)}
\label{homo}
\vspace{-2em}
\begin{center}
$$
\begin{array}{c | c c c c | c}
       & \phantom{\Big|}                        1 &                     2
       & \cdots &                         s & \\\hline
     1 & \phantom{\Big|}n_{1,\s} \cdot n_{\s,1}/n & n_{1,\s} \cdot n_{\s,2}/n
       & \cdots & n_{1,\s} \cdot n_{\s,s} / n & n_{1,\s} \\
     2 & \phantom{\Big|}n_{2,\s} \cdot n_{\s,1}/n & n_{2,\s} \cdot n_{\s,2}/n
       & \cdots & n_{2,\s} \cdot n_{\s,s} / n & n_{2,\s} \\
\vdots & \phantom{\Big|}                   \vdots &                    \vdots
       & \ddots &                    \vdots &  \vdots \\
     r & \phantom{\Big|}n_{r,\s} \cdot n_{\s,1}/n & n_{r,\s} \cdot n_{\s,2}/n
       & \cdots & n_{r,\s} \cdot n_{\s,s} / n & n_{r,\s} \\\hline
       & \phantom{\Big|}                  n_{\s,1} &                   n_{\s,2}
       & \cdots &                   n_{\s,s} & n_{\s,\s}
\end{array}
$$
\end{center}
\end{table}

The statistical analysis of categorical data is commonly formulated
in the framework of contingency-tables/cross-tabulations;
Table~\ref{tab} provides a typical two-way example
\citep[see, for instance, Chapter~4 of]
      [for a comprehensive treatment]{andersen}.
A common task is to ascertain whether the given data
(displayed in Table~\ref{tab}) is consistent
up to expected statistical fluctuations
with the model for homogeneity of proportions (displayed in Table~\ref{homo}).
When considering homogeneity, we assume that the probabilistic process
generating the given data fixes the column totals (but not the row totals)
by construction.
Therefore, to gauge whether the given data displayed in Table~\ref{tab}
is consistent with the assumed homogeneity displayed
in Table~\ref{homo}, we do the following:
\begin{enumerate}
\item We generate $s$ sets of draws,
with the $k$th set consisting of $n_{\s,k}$ independent and identically
distributed draws from the probability distribution $(p_1, p_2, \dots, p_r)$,
where $p_1 = n_{1,\s}/n$, $p_2 = n_{2,\s}/n$, \dots, $p_r = n_{r,\s}/n$.
Note that $p_j = (n_{j,\s} \cdot n_{\s,k} / n) / n_{\s,k}$
for $j = 1$,~$2$, \dots, $r$;
these are homogeneous proportions (since $p_1$,~$p_2$, \dots, $p_r$
are the same for every column index $k$).
\item For each of the $s$ sets of draws --- say the $k$th set ---
we define $N_{j,k}$ to be the number of draws falling in the $j$th row,
for $j = 1$,~$2$, \dots, $r$.
\item We calculate the probability $P$ that the discrepancy
between the simulated counts $N_{j,k}$ and the model
$N_{j,\s} \cdot N_{\s,k} / n$ is greater than or equal to the discrepancy
between the observed counts $n_{j,k}$ and the assumed
$n_{j,\s} \cdot n_{\s,k} / n$. When calculating this probability,
we view $N_{j,k}$ and $N_{j,\s}$ as random,
while viewing all other numbers as fixed.
Please note that, by construction,
$N_{\s,k} = n_{\s,k}$ for $k = 1$,~$2$, \dots, $s$.
\end{enumerate}

The number $P$ defined in Step~3 is known as the (exact) P-value.
Given the P-value $P$, we can have $100(1-P)\%$ confidence
that the observed draws are not consistent
with assuming the homogeneity displayed in Table~\ref{homo}.
See Section~3 of~\cite{perkins-tygert-ward3} for further discussion
of P-values and their interpretation;
Section~3 of~\cite{perkins-tygert-ward3} details subtleties involved
in the definition and interpretation of these P-values.

The definition above of the P-value $P$ requires a metric
for measuring the discrepancies. The canonical choices are $\chi^2$
and the log--likelihood-ratio $G^2$:

\begin{equation}
\chi^2 = \sum_{j=1}^r \sum_{k=1}^s
\frac{(n_{j,k} - (n_{j,\s} \cdot n_{\s,k}/n))^2}{n_{j,\s} \cdot n_{\s,k}/n}
\end{equation}
\begin{equation}
X^2 = \sum_{j=1}^r \sum_{k=1}^s
\frac{(N_{j,k} - (N_{j,\s} \cdot N_{\s,k}/n))^2}{N_{j,\s} \cdot N_{\s,k}/n}
\end{equation}
\begin{equation}
\label{pchi2}
P_{\chi^2} = \Prob\{X^2 \ge \chi^2\}
\end{equation}

\nopagebreak
\begin{equation}
g^2 = 2 \sum_{j=1}^r \sum_{k=1}^s n_{j,k}
        \cdot \ln\left(\frac{n_{j,k}}{n_{j,\s} \cdot n_{\s,k}/n}\right)
\end{equation}
\begin{equation}
G^2 = 2 \sum_{j=1}^r \sum_{k=1}^s N_{j,k}
        \cdot \ln\left(\frac{N_{j,k}}{N_{j,\s} \cdot N_{\s,k}/n}\right)
\end{equation}
\begin{equation}
\label{pg2}
P_{g^2} = \Prob\{G^2 \ge g^2\}
\end{equation}

\noindent Other possibilities include the Hellinger (or Freeman-Tukey) distance
and the Frobenius (or Hilbert-Schmidt or Euclidean) distance:

\begin{equation}
h^2 = 4 \sum_{j=1}^r \sum_{k=1}^s
(\sqrt{n_{j,k}} - \sqrt{n_{j,\s} \cdot n_{\s,k}/n})^2
\end{equation}
\begin{equation}
H^2 = 4 \sum_{j=1}^r \sum_{k=1}^s
(\sqrt{N_{j,k}} - \sqrt{N_{j,\s} \cdot N_{\s,k}/n})^2
\end{equation}
\begin{equation}
\label{ph2}
P_{h^2} = \Prob\{H^2 \ge h^2\}
\end{equation}

\begin{equation}
\label{deff2}
f^2 = \sum_{j=1}^r \sum_{k=1}^s (n_{j,k} - (n_{j,\s} \cdot n_{\s,k}/n))^2
\end{equation}
\begin{equation}
F^2 = \sum_{j=1}^r \sum_{k=1}^s (N_{j,k} - (N_{j,\s} \cdot N_{\s,k}/n))^2
\end{equation}
\begin{equation}
\label{pf2}
P_{f^2} = \Prob\{F^2 \ge f^2\}
\end{equation}

\noindent When taking probabilities in~(\ref{pchi2}), (\ref{pg2}),
(\ref{ph2}), and~(\ref{pf2}), we view the uppercase $X^2$, $G^2$, $H^2$,
and $F^2$ as random variables, while viewing the lowercase $\chi^2$, $g^2$,
$h^2$, and $f^2$ as fixed numbers.

As discussed, for example, by~\cite{rao}, $X^2$, $G^2$, and $H^2$ all converge
to the same distribution in the limit of large numbers of draws ---
$X^2$, $G^2$, and $H^2$ are the best-known members
of the Cressie-Read power-divergence family.
$F^2$ is not a member of the Cressie-Read power-divergence family
and does not necessarily converge to the same distribution
as $X^2$, $G^2$, and $H^2$.
\cite{perkins-tygert-ward3} illustrated the many advantages of $F^2$
when neither the row totals nor the column totals are fixed;
the present paper illustrates the advantages when the column totals are fixed.
However, $F^2$ is not uniformly more powerful than the classical statistics.
We recommend using both $F^2$ and a classical statistic such as $G^2$.

In the sequel, Section~\ref{computation} summarizes an algorithm
for computing the P-values defined above.
Section~\ref{dataa} analyzes several data sets.
Section~\ref{conclusion} draws some conclusions.

\section{Computation of P-values}
\label{computation}

The definitions of the P-values in~(\ref{pchi2}), (\ref{pg2}), (\ref{ph2}),
and~(\ref{pf2}) involve the probabilities of certain events.
In the present paper, we compute these probabilities
via Monte-Carlo simulations with guaranteed error bounds.
Specifically, we conduct a large number $m$ of simulations;
in each simulation --- say the $\ell$th ---
we perform the following steps (using the data of Table~\ref{tab}):
\begin{enumerate}
\item We generate $s$ sets of draws,
with the $k$th set consisting of $n_{\s,k}$ independent and identically
distributed draws from the probability distribution $(p_1, p_2, \dots, p_r)$,
where $p_1 = n_{1,\s}/n$, $p_2 = n_{2,\s}/n$, \dots, $p_r = n_{r,\s}/n$.
Note that $p_j = (n_{j,\s} \cdot n_{\s,k} / n) / n_{\s,k}$
for $j = 1$,~$2$, \dots, $r$;
these are homogeneous proportions (since $p_1$,~$p_2$, \dots, $p_r$
are the same for every column index $k$).
Furthermore, the underlying distribution of the draws does not depend
on $\ell$.
\item For each of the $s$ sets of draws --- say the $k$th set ---
we define $n^{(\ell)}_{j,k}$ to be the number of draws falling
in the $j$th row, for $j = 1$,~$2$, \dots, $r$.
\item We calculate the discrepancy $f^2_{(\ell)}$
between the simulated counts $n^{(\ell)}_{j,k}$ and the model
$n^{(\ell)}_{j,\s} \cdot n^{(\ell)}_{\s,k} / n$, that is,
\begin{equation}
f_{(\ell)}^2 = \sum_{j=1}^r \sum_{k=1}^s
(n^{(\ell)}_{j,k} - (n^{(\ell)}_{j,\s} \cdot n^{(\ell)}_{\s,k}/n))^2.
\end{equation}
\end{enumerate}

An estimate of the P-value $P_{f^2}$ is the fraction
of $f_{(1)}^2$,~$f_{(2)}^2$, \dots, $f_{(m)}^2$ which are greater than
or equal to $f^2$ defined in~(\ref{deff2}).
As discussed in Section~3 of~\cite{perkins-tygert-ward3},
the standard error of the estimate is $\sqrt{P_{f^2}(1-P_{f^2})/m}$,
where $m$ is the number of simulations.

Needless to say, we can compute the P-values for $\chi^2$, $g^2$, and $h^2$
via similar procedures, with the same error bounds.

\begin{remark}
For all computations reported in the present paper,
we generated random numbers via the C programming language procedure
given on page~9 of~\cite{marsaglia},
implementing the recommended complementary multiply with carry.
\end{remark}

\section{Data analysis}
\label{dataa}

To compare the performance of the various metrics for measuring
the discrepancies between observed and simulated data,
we analyze several data sets.
Using the procedure of Section~\ref{computation},
we conduct $m =$ 4,000,000 Monte-Carlo simulations per P-value,
for each of the examples presented below.
The standard error of the resulting estimate for the P-value $P$
is then $\sqrt{P(1-P)}/2000$; see Section~3 of~\cite{perkins-tygert-ward3}.
Before reporting the P-values associated with the data sets,
we make two remarks concerning their interpretation:

\begin{remark}
A significance test can only indicate that observed data
{\it cannot} be reasonably assumed to have arisen from the model
of homogeneous proportions; a significance test cannot prove
that the observed data {\it can} be reasonably assumed to have arisen
from the model of homogeneity.
Thus, aside from considerations of multiple testing,
if any statistic strongly signals that the data cannot be reasonably assumed
to have arisen from the model of homogeneity, then we must reject
(or at least question) the model --- irrespective of any large P-values
for other statistics. For instance, if the P-value for the Frobenius distance
$f^2$ is very small, then we should not accept the model of homogeneity,
not even if the P-values for $\chi^2$, the log--likelihood-ratio $g^2$,
and the Freeman-Tukey/Hellinger-distance $h^2$ are large.
\end{remark}

\begin{remark}
\label{nll}
The term ``negative log-likelihood'' used in the present section refers
to the statistic that is simply the negative of the logarithm
of the likelihood.
The negative log-likelihood is the same statistic used
in the generalization of Fisher's exact test
discussed by~\cite{guo-thompson}; unlike the log--likelihood-ratio $G^2$,
this statistic involves only one likelihood, not the ratio of two.
We mention the negative log-likelihood just to facilitate comparisons;
we are not asserting that the likelihood
on its own (rather than in a ratio) is a good gauge
of the relative sizes of deviations from a model.
\end{remark}

The $11 \times 2$ Table~\ref{Danish} displays the data for our first example,
which has 22 entries in all.
Table~\ref{Danishh} displays the model of homogeneous proportions
for Table~\ref{Danish}.
The P-values for Table~\ref{Danish} for the assumption that Table~\ref{Danishh}
gives the correct underlying distribution are

\medskip
\centerline{
\begin{tabular}{rl}
$\chi^2$ ($X^2$): & .0868 \\
log--likelihood-ratio ($G^2$): & .0906 \\
Freeman-Tukey/Hellinger ($H^2$): & .0959 \\
negative log-likelihood: & .0905 \\
Frobenius ($F^2$): & .00838
\end{tabular}
}\medskip

\noindent Please note that the P-value for the Frobenius distance
is over an order of magnitude smaller than the P-values
for the classical statistics. 

The $7 \times 3$ Table~\ref{mania} displays the data for our second example,
which has 21 entries in all.
Table~\ref{maniah} displays the model of homogeneous proportions
for Table~\ref{mania}.
The P-values for Table~\ref{mania} for the assumption that Table~\ref{maniah}
gives the correct underlying distribution are

\medskip
\centerline{
\begin{tabular}{rl}
$\chi^2$ ($X^2$): & .145 \\
log--likelihood-ratio ($G^2$): & .292 \\
Freeman-Tukey/Hellinger ($H^2$): & .493 \\
negative log-likelihood: & .132 \\
Frobenius ($F^2$): & .0286
\end{tabular}
}\medskip

\noindent Please note that the P-value for the Frobenius distance is
over four times smaller than the P-values for the classical statistics. 

The $9 \times 2$ Table~\ref{Republican} displays the data
for our third example, which has 18 entries in all.
Table~\ref{Republicanh} displays the model of homogeneous proportions
for Table~\ref{Republican}.
The P-values for Table~\ref{Republican} for the assumption
that Table~\ref{Republicanh} gives the correct underlying distribution are

\medskip
\centerline{
\begin{tabular}{rl}
$\chi^2$ ($X^2$): & .123 \\
log--likelihood-ratio ($G^2$): & .138 \\
Freeman-Tukey/Hellinger ($H^2$): & .157 \\
negative log-likelihood: & .114 \\
Frobenius ($F^2$): & .0344
\end{tabular}
}\medskip

\noindent Please note that the P-value for the Frobenius distance is
over three times smaller than the P-values for the classical statistics. 

The $5 \times 3$ Table~\ref{mania2} displays the data for our final example,
which has 15 entries in all.
Table~\ref{mania2h} displays the model of homogeneous proportions
for Table~\ref{mania2}.
The P-values for Table~\ref{mania2} for the assumption that Table~\ref{mania2h}
gives the correct underlying distribution are

\medskip
\centerline{
\begin{tabular}{rl}
$\chi^2$ ($X^2$): & .276 \\
log--likelihood-ratio ($G^2$): & .171 \\
Freeman-Tukey/Hellinger ($H^2$): & .0794 \\
negative log-likelihood: & .235 \\
Frobenius ($F^2$): & .199
\end{tabular}
}\medskip

\noindent In this example, none of the statistics produces
a very small P-value; the smallest arises
from the Freeman-Tukey/Hellinger distance in this case.

\begin{remark}
Appropriate binning (or rebinning) to uniformize the frequencies associated
with the entries in the contingency-tables/cross-tabulations
can mitigate the problem with the classical statistics.
Yet rebinning is a black art that is liable to improperly influence the result
of a significance test, and the usual data-dependent rebinning calls
for Monte-Carlo simulations to calculate P-values accurately anyways.
Rebinning always requires careful extra work.
A principal advantage of the Frobenius distance
is that it does not require any rebinning; indeed, the Frobenius distance
is most powerful without any rebinning.
Note also that optimally rebinning data such as that displayed
in Table~\ref{Danish} can be very challenging.
\end{remark}

\begin{table}[p]
\caption{Results of polls in June 1983 for Danish parliamentary elections,
from Chapter~4 of~\cite{andersen}}
\label{Danish}
\begin{center}
\begin{tabular}{lrrrrrrrr}
Party  &&&  Poll 1 &           &&& Poll 2 &           \\\hline
     A &&&     416 &  (33.1\%) &&&    268 &  (38.9\%) \\
     B &&&      45 &   (3.6\%) &&&     22 &   (3.2\%) \\
     C &&&     338 &  (26.9\%) &&&    160 &  (23.2\%) \\
     E &&&      13 &   (1.0\%) &&&      6 &   (0.9\%) \\
     F &&&     131 &  (10.4\%) &&&     66 &   (9.6\%) \\
     K &&&      18 &   (1.4\%) &&&     10 &   (1.5\%) \\
     M &&&      47 &   (3.7\%) &&&     16 &   (2.3\%) \\
     Q &&&      20 &   (1.6\%) &&&      8 &   (1.2\%) \\
     V &&&     129 &  (10.3\%) &&&     92 &  (13.4\%) \\
     Y &&&      22 &   (1.8\%) &&&      9 &   (1.3\%) \\
     Z &&&      76 &   (6.1\%) &&&     32 &   (4.6\%) \\\hline
   All &&&    1255 & (100.0\%) &&&    689 & (100.0\%)
\end{tabular}
\end{center}
\end{table}

\begin{table}
\caption{The model of homogeneous proportions for Table~\ref{Danish}}
\label{Danishh}
\begin{center}
\begin{tabular}{lrrrrrrrr}
Party  &&&  Poll 1 &           &&& Poll 2 &           \\\hline
     A &&&   441.6 &  (35.2\%) &&&  242.4 &  (35.2\%) \\
     B &&&    43.3 &   (3.4\%) &&&   23.7 &   (3.4\%) \\
     C &&&   321.5 &  (25.6\%) &&&  176.5 &  (25.6\%) \\
     E &&&    12.3 &   (1.0\%) &&&    6.7 &   (1.0\%) \\
     F &&&   127.2 &  (10.1\%) &&&   69.8 &  (10.1\%) \\
     K &&&    18.1 &   (1.4\%) &&&    9.9 &   (1.4\%) \\
     M &&&    40.7 &   (3.2\%) &&&   22.3 &   (3.2\%) \\
     Q &&&    18.1 &   (1.4\%) &&&    9.9 &   (1.4\%) \\
     V &&&   142.7 &  (11.4\%) &&&   78.3 &  (11.4\%) \\
     Y &&&    20.0 &   (1.6\%) &&&   11.0 &   (1.6\%) \\
     Z &&&    69.7 &   (5.6\%) &&&   38.3 &   (5.6\%) \\\hline
   All &&&  1255.0 & (100.0\%) &&&  689.0 & (100.0\%) 
\end{tabular}
\end{center}
\end{table}

\begin{table}
\caption{Differences between the entries of Table~\ref{Danish}
         and the corresponding entries of Table~\ref{Danishh}}
\label{Danishhd}
\begin{center}
\begin{tabular}{lrr}
Party  &  Poll 1 & Poll 2 \\\hline
     A & $-25.6$ & $ 25.6$ \\
     B & $  1.7$ & $ -1.7$ \\
     C & $ 16.5$ & $-16.5$ \\
     E & $  0.7$ & $ -0.7$ \\
     F & $  3.8$ & $ -3.8$ \\
     K & $ -0.1$ & $  0.1$ \\
     M & $  6.3$ & $ -6.3$ \\
     Q & $  1.9$ & $ -1.9$ \\
     V & $-13.7$ & $ 13.7$ \\
     Y & $  2.0$ & $ -2.0$ \\
     Z & $  6.3$ & $ -6.3$ \\\hline
   All & $  0.0$ & $  0.0$
\end{tabular}
\end{center}
\end{table}

\begin{table}
\caption{The entries of Table~\ref{Danishhd} divided by the square roots
         of the corresponding entries of Table~\ref{Danishh}}
\label{Danishhdn}
\begin{center}
\begin{tabular}{lrr}
Party  & Poll 1 & Poll 2 \\\hline
     A & $-1.2$ & $ 1.6$ \\
     B & $ 0.3$ & $-0.4$ \\
     C & $ 0.9$ & $-1.2$ \\
     E & $ 0.2$ & $-0.3$ \\
     F & $ 0.3$ & $-0.5$ \\
     K & $-0.0$ & $ 0.0$ \\
     M & $ 1.0$ & $-1.3$ \\
     Q & $ 0.5$ & $-0.6$ \\
     V & $-1.1$ & $ 1.5$ \\
     Y & $ 0.4$ & $-0.6$ \\
     Z & $ 0.8$ & $-1.0$ \\\hline
   All & $ 0.0$ & $ 0.0$
\end{tabular}
\end{center}
\end{table}

\begin{table}
\caption{Reasons for (or absence of) premature termination of the treatment
of maniacal patients in three groups from~\cite{bowden-et_al}
(the three groups are those treated with divalproex,
those treated with lithium, and those ``treated'' with a placebo)}
\label{mania}
\begin{center}
\begin{tabular}{lrrrrrrrr}
          Reason & Divalproex & && Lithium & && Placebo & \\\hline
Lack of efficacy & 21 &  (30.4\%) && 12 &  (33.3\%) && 38 &  (51.4\%) \\
     Intolerance &  4 &   (5.8\%) &&  4 &  (11.1\%) &&  2 &   (2.7\%) \\
       Recovered &  3 &   (4.3\%) &&  2 &   (5.6\%) &&  2 &   (2.7\%) \\
   Noncompliance &  1 &   (1.4\%) &&  1 &   (2.8\%) &&  3 &   (4.1\%) \\
 Another illness &  0 &   (0.0\%) &&  1 &   (2.8\%) &&  0 &   (0.0\%) \\
  Administration &  4 &   (5.8\%) &&  2 &   (5.6\%) &&  2 &   (2.7\%) \\
  Not terminated & 36 &  (52.2\%) && 14 &  (38.9\%) && 27 &  (36.5\%) \\\hline
             All & 69 & (100.0\%) && 36 & (100.0\%) && 74 & (100.0\%)
\end{tabular}
\end{center}
\end{table}

\begin{table}
\caption{The model of homogeneous proportions for Table~\ref{mania}}
\label{maniah}
\begin{center}
\begin{tabular}{lrrrrrrrr}
          Reason & Divalproex & && Lithium &  && Placebo & \\\hline
Lack of efficacy & 27.4 &  (39.7\%) && 14.3 &  (39.7\%) && 29.4 &  (39.7\%) \\
     Intolerance &  3.9 &   (5.6\%) &&  2.0 &   (5.6\%) &&  4.1 &   (5.6\%) \\
       Recovered &  2.7 &   (3.9\%) &&  1.4 &   (3.9\%) &&  2.9 &   (3.9\%) \\
   Noncompliance &  1.9 &   (2.8\%) &&  1.0 &   (2.8\%) &&  2.1 &   (2.8\%) \\
 Another illness &  0.4 &   (0.6\%) &&  0.2 &   (0.6\%) &&  0.4 &   (0.6\%) \\
  Administration &  3.1 &   (4.5\%) &&  1.6 &   (4.5\%) &&  3.3 &   (4.5\%) \\
  Not terminated & 29.7 &  (43.0\%) && 15.5 &  (43.0\%) && 31.8 &  (43.0\%) \\
                                                                   \hline
             All & 69.0 & (100.0\%) && 36.0 & (100.0\%) && 74.0 & (100.0\%)
\end{tabular}
\end{center}
\end{table}

\begin{table}
\caption{Differences between the entries of Table~\ref{mania}
         and the corresponding entries of Table~\ref{maniah}}
\label{maniahd}
\begin{center}
\begin{tabular}{lrrr}
          Reason & Divalproex & Lithium & Placebo \\\hline
Lack of efficacy &     $-6.4$ &  $-2.3$ &   $8.6$ \\
     Intolerance &      $0.1$ &   $2.0$ &  $-2.1$ \\
       Recovered &      $0.3$ &   $0.6$ &  $-0.9$ \\
   Noncompliance &     $-0.9$ &   $0.0$ &   $0.9$ \\
 Another illness &     $-0.4$ &   $0.8$ &  $-0.4$ \\
  Administration &      $0.9$ &   $0.4$ &  $-1.3$ \\
  Not terminated &      $6.3$ &  $-1.5$ &  $-4.8$ \\\hline
             All &      $0.0$ &   $0.0$ &   $0.0$
\end{tabular}
\end{center}
\end{table}

\begin{table}
\caption{The entries of Table~\ref{maniahd} divided by the square roots
         of the corresponding entries of Table~\ref{maniah}}
\label{maniahdn}
\begin{center}
\begin{tabular}{lrrr}
          Reason & Divalproex & Lithium & Placebo \\\hline
Lack of efficacy &     $-1.2$ &  $-0.6$ &  $ 1.6$ \\
     Intolerance &     $ 0.1$ &  $ 1.4$ &  $-1.0$ \\
       Recovered &     $ 0.2$ &  $ 0.5$ &  $-0.5$ \\
   Noncompliance &     $-0.7$ &  $ 0.0$ &  $ 0.6$ \\
 Another illness &     $-0.6$ &  $ 1.8$ &  $-0.6$ \\
  Administration &     $ 0.5$ &  $ 0.3$ &  $-0.7$ \\
  Not terminated &     $ 1.2$ &  $-0.4$ &  $-0.9$ \\\hline
             All &     $ 0.0$ &  $ 0.0$ &  $ 0.0$
\end{tabular}
\end{center}
\end{table}

\begin{table}
\caption{Results for the 2012 Republican U.S. presidential nomination,
from a CBS News poll of November 6--10, 2011 (released November 11, 2011)
and from a Pew Research Center poll of November 9--11, 2011
(released November 17, 2011), as reconstructed from percentages
rounded to the nearest whole numbers (the original counts were not reported)
for Republican primary voters}
\label{Republican}
\begin{center}
\begin{tabular}{lrrrrrrrrrr}
       Candidate &&&& CBS &           &&&& Pew &           \\\hline
Michele Bachmann &&&&  15 &   (4.6\%) &&&&  21 &   (5.1\%) \\
     Herman Cain &&&&  69 &  (21.2\%) &&&& 103 &  (25.0\%) \\
   Newt Gingrich &&&&  57 &  (17.5\%) &&&&  66 &  (16.0\%) \\
    Jon Huntsman &&&&   4 &   (1.2\%) &&&&   4 &   (1.0\%) \\
        Ron Paul &&&&  19 &   (5.8\%) &&&&  33 &   (8.0\%) \\
      Rick Perry &&&&  31 &   (9.5\%) &&&&  37 &   (9.0\%) \\
     Mitt Romney &&&&  57 &  (17.5\%) &&&&  91 &  (22.1\%) \\
   Rick Santorum &&&&   8 &   (2.5\%) &&&&   8 &   (1.9\%) \\
     Do not know &&&&  65 &  (20.0\%) &&&&  49 &  (11.9\%) \\\hline
             All &&&& 325 & (100.0\%) &&&& 412 & (100.0\%)
\end{tabular}
\end{center}
\end{table}

\begin{table}
\caption{The model of homogeneous proportions for Table~\ref{Republican}}
\label{Republicanh}
\begin{center}
\begin{tabular}{lrrrrrrrrrr}
       Candidate &&&&   CBS &           &&&&   Pew &           \\\hline
Michele Bachmann &&&&  15.9 &   (4.9\%) &&&&  20.1 &   (4.9\%) \\
     Herman Cain &&&&  75.8 &  (23.3\%) &&&&  96.2 &  (23.3\%) \\
   Newt Gingrich &&&&  54.2 &  (16.7\%) &&&&  68.8 &  (16.7\%) \\
    Jon Huntsman &&&&   3.5 &   (1.1\%) &&&&   4.5 &   (1.1\%) \\
        Ron Paul &&&&  22.9 &   (7.1\%) &&&&  29.1 &   (7.1\%) \\
      Rick Perry &&&&  30.0 &   (9.2\%) &&&&  38.0 &   (9.2\%) \\
     Mitt Romney &&&&  65.3 &  (20.1\%) &&&&  82.7 &  (20.1\%) \\
   Rick Santorum &&&&   7.1 &   (2.2\%) &&&&   8.9 &   (2.2\%) \\
     Do not know &&&&  50.3 &  (15.5\%) &&&&  63.7 &  (15.5\%) \\\hline
             All &&&& 325.0 & (100.0\%) &&&& 412.0 & (100.0\%)
\end{tabular}
\end{center}
\end{table}

\begin{table}
\caption{Differences between the entries of Table~\ref{Republican}
         and the corresponding entries of Table~\ref{Republicanh}}
\label{Republicanhd}
\begin{center}
\begin{tabular}{lrr}
       Candidate &    CBS &     Pew \\\hline
Michele Bachmann & $-0.9$ &   $0.9$ \\
     Herman Cain & $-6.8$ &   $6.8$ \\
   Newt Gingrich &  $2.8$ &  $-2.8$ \\
    Jon Huntsman &  $0.5$ &  $-0.5$ \\
        Ron Paul & $-3.9$ &   $3.9$ \\
      Rick Perry &  $1.0$ &  $-1.0$ \\
     Mitt Romney & $-8.3$ &   $8.3$ \\
   Rick Santorum &  $0.9$ &  $-0.9$ \\
     Do not know & $14.7$ & $-14.7$ \\\hline
             All &  $0.0$ &   $0.0$
\end{tabular}
\end{center}
\end{table}

\begin{table}
\caption{The entries of Table~\ref{Republicanhd} divided by the square roots
         of the corresponding entries of Table~\ref{Republicanh}}
\label{Republicanhdn}
\begin{center}
\begin{tabular}{lrr}
       Candidate &    CBS &    Pew \\\hline
Michele Bachmann & $-0.2$ & $ 0.2$ \\
     Herman Cain & $-0.8$ & $ 0.7$ \\
   Newt Gingrich & $ 0.4$ & $-0.3$ \\
    Jon Huntsman & $ 0.3$ & $-0.2$ \\
        Ron Paul & $-0.8$ & $ 0.7$ \\
      Rick Perry & $ 0.2$ & $-0.2$ \\
     Mitt Romney & $-1.0$ & $ 0.9$ \\
   Rick Santorum & $ 0.4$ & $-0.3$ \\
     Do not know & $ 2.1$ & $-1.8$ \\\hline
             All & $ 0.0$ & $ 0.0$
\end{tabular}
\end{center}
\end{table}

\begin{table}
\caption{Reactions to prior treatment with lithium
(when treated before with lithium)
of maniacal patients in three groups from~\cite{bowden-et_al}
(the three groups are those treated with divalproex,
those treated with lithium, and those ``treated'' with a placebo)}
\label{mania2}
\vspace{.5em}
\begin{center}
\begin{tabular}{lrrrrrrrr}
Reaction
& Divalproex && & Lithium && & Placebo & \\\\\hline\\
\parbox[c]{1.2in}{Effective and\\tolerated}
& 22 &  (31.9\%) && 16 &  (44.4\%) && 19 &  (25.7\%) \\\\
\parbox[c]{1.2in}{Effective but\\not tolerated}
&  7 &  (10.1\%) &&  0 &   (0.0\%) &&  6 &   (8.1\%) \\\\
\parbox[c]{1.2in}{Ineffective but\\tolerated}
& 19 &  (27.5\%) && 11 &  (30.6\%) && 31 &  (41.9\%) \\\\
\parbox[c]{1.2in}{Ineffective and\\not tolerated}
&  6 &   (8.7\%) &&  4 &  (11.1\%) &&  5 &   (6.8\%) \\\\
\parbox[c]{1.2in}{No prior lithium\\treatment}
& 15 &  (21.7\%) &&  5 &  (13.9\%) && 13 &  (17.6\%) \\\\\hline\\
All
& 69 & (100.0\%) && 36 & (100.0\%) && 74 & (100.0\%)
\end{tabular}
\end{center}
\end{table}

\begin{table}
\caption{The model of homogeneous proportions for Table~\ref{mania2}}
\label{mania2h}
\vspace{.5em}
\begin{center}
\begin{tabular}{lrrrrrrrr}
Reaction
& Divalproex && & Lithium && & Placebo & \\\\\hline\\
\parbox[c]{1.2in}{Effective and\\tolerated}
& 22.0 &  (31.8\%) && 11.5 &  (31.8\%) && 23.6 &  (31.8\%) \\\\
\parbox[c]{1.2in}{Effective but\\not tolerated}
&  5.0 &   (7.3\%) &&  2.6 &   (7.3\%) &&  5.4 &   (7.3\%) \\\\
\parbox[c]{1.2in}{Ineffective but\\tolerated}
& 23.5 &  (34.1\%) && 12.3 &  (34.1\%) && 25.2 &  (34.1\%) \\\\
\parbox[c]{1.2in}{Ineffective and\\not tolerated}
&  5.8 &   (8.4\%) &&  3.0 &   (8.4\%) &&  6.2 &   (8.4\%) \\\\
\parbox[c]{1.2in}{No prior lithium\\treatment}
& 12.7 &  (18.4\%) &&  6.6 &  (18.4\%) && 13.6 &  (18.4\%) \\\\\hline\\
All
& 69.0 & (100.0\%) && 36.0 & (100.0\%) && 74.0 & (100.0\%)
\end{tabular}
\end{center}
\end{table}

\begin{table}
\caption{Differences between the entries of Table~\ref{mania2}
         and the corresponding entries of Table~\ref{mania2h}}
\label{mania2hd}
\begin{center}
\begin{tabular}{lrrr}
                     Reaction & Divalproex & Lithium & Placebo \\\hline
      Effective and tolerated &      $0.0$ &   $4.5$ &  $-4.6$ \\
  Effective but not tolerated &      $2.0$ &  $-2.6$ &   $0.6$ \\
    Ineffective but tolerated &     $-4.5$ &  $-1.3$ &   $5.8$ \\
Ineffective and not tolerated &      $0.2$ &   $1.0$ &  $-1.2$ \\
   No prior lithium treatment &      $2.3$ &  $-1.6$ &  $-0.6$ \\\hline
                          All &      $0.0$ &   $0.0$ &   $0.0$
\end{tabular}
\end{center}
\end{table}

\begin{table}
\caption{The entries of Table~\ref{mania2hd} divided by the square roots
         of the corresponding entries of Table~\ref{mania2h}}
\label{mania2hdn}
\begin{center}
\begin{tabular}{lrrr}
                     Reaction & Divalproex & Lithium & Placebo \\\hline
      Effective and tolerated &     $ 0.0$ &  $ 1.3$ & $-0.9$ \\
  Effective but not tolerated &     $ 0.9$ &  $-1.6$ & $ 0.3$ \\
    Ineffective but tolerated &     $-0.9$ &  $-0.4$ & $ 1.2$ \\
Ineffective and not tolerated &     $ 0.1$ &  $ 0.6$ & $-0.5$ \\
   No prior lithium treatment &     $ 0.6$ &  $-0.6$ & $-0.2$ \\\hline
                          All &     $ 0.0$ &  $ 0.0$ & $ 0.0$
\end{tabular}
\end{center}
\end{table}

\section{Conclusion}
\label{conclusion}

The Frobenius distance is significantly more powerful
than the classical statistics for gauging the consistency
of observed data with the assumption of homogeneity
in many of the examples of the present paper.
This may or may not be typical of most applications;
actually, we suspect that our last example ---
in which all the statistics perform similarly --- is the most representative.
Even so, both the present paper and the applications
of~\cite{perkins-tygert-ward3} illustrate that
there are many important circumstances in which the Frobenius distance
is much more powerful than the classical statistics.

\section*{Acknowledgements}

We would like to thank Alex Barnett, G\'erard Ben Arous, James Berger,
Tony Cai, Sourav Chatterjee, Ronald Raphael Coifman, Ingrid Daubechies,
Jianqing Fan, Jiayang Gao, Andrew Gelman, Leslie Greengard, Peter W. Jones,
Deborah Mayo, Peter McCullagh, Michael O'Neil, Ron Peled, William Perkins,
William H. Press, Vladimir Rokhlin, Joseph Romano, Gary Simon, Amit Singer,
Michael Stein, Stephen Stigler, Joel Tropp, Rachel Ward, Larry Wasserman,
and Douglas A. Wolfe.
This work was supported in part by a research fellowship
from the Alfred P. Sloan Foundation.

\newpage

\addcontentsline{toc}{section}{\protect\numberline{}References}
\bibliographystyle{asamod.bst}
\bibliography{stat}

\end{document}